# Vacuum Acceptance Tests for Particle Accelerator Equipment

*G.Bregliozzi*
CERN, Geneva, Switzerland

**Abstract**

The effective and reliable operation of particle accelerator machines is strongly related to obtaining and keeping the required ultra-high vacuum level. This paper briefly presents some generic requirements to take in consideration for components that need to be installed in a particle accelerator like design constraints, materials choices, chemical cleaning process, outgassing rate, and their possible mitigation. Moreover, it will discuss vacuum acceptance tests put in place in the CERN accelerator complex for both, baked and unbaked system.

**Keywords**
Acceptance requirements, residual gas analysis, outgassing rate, virtual leak

## 1    Introduction

In a vacuum system, contamination might be defined as anything which could prevent the vacuum system from reaching the desired base pressure, introduce unwanted or detrimental species into the residual gas or that could, to some extent, modify the surface properties of all or part of it. Experience has shown that in order to achieve stringent vacuum requirements, chemical cleaning processes and materials selection are fundamental steps. In general, each vessel and components have been machined and worked in some way. Such process could use oil which, having a high outgassing rate or vapour pressure, could limit the base pressure. Moreover, water, solvent and other liquids can remain embedded in cracks and pores and can outgas over a long period. Presence of polymeric materials in a vacuum system, due to their elevate outgassing, could induce localized high pressure and consequently decreasing considerably the lifetime of pumping system like the sputter ionic pumps leading, also in this case, to limited performance and premature failure [1]. In this case, we could talk about contamination due to the high outgassing rate of the chosen material. Another example could be a system operating at $10^{-8}$ mbar, in presence of $10^{-10}$ mbar partial pressure of hydrocarbon; these molecules could strike a mirror surface and polymerize when irradiated by photon leading to graphite-like or insulating layer. This leads to altered optical properties of the mirror and thus we could define this partial pressure as a source of contamination [2-3]. In addition to these examples, it is important to also stress out that one of the most important lifetime limits in storage rings is beam gas scattering. Beam scattering increases with the square of the atomic number Z of the gas species present in the vacuum beam pipe. Thus, it is important to minimize as much as possible the partial pressure of high-Z species, otherwise, these should be considered as contaminants of the system [4].

For these reasons and much more, it is clear that it is the role of the vacuum expert to provide guidance and on-going support to ensure the implementation of compatible vacuum equipment. This responsibility extends to supporting these systems during commissioning and operations, too. To achieve that mandate, acceptance tests on partial systems and fully assembly are essential, which include the measurement of leak tightness, residual gas analysis, level of contamination, outgassing rate and its time variation and virtual leaks (inleakage).

In the following paragraphs, an overview of the acceptance criteria used at CERN for unbaked and baked system will be presented. It is important to point out that this paper is not a guideline but it is just a summary of gained experience over different years at CERN. It also highlights the role of the vacuum expert in the definition of components for different vacuum systems as it is presented in different reports and functional specifications of different storage rings and accelerator complex worldwide [5-13].

## 2 Specification of vacuum system quality

Two main aspects should be kept in mind when defining the quality and consequently the acceptance criteria of a vacuum system: specification of the vacuum system by the process and specification of the vacuum system by its performance.

### 2.1 How to define the right specification

The vast majority of users of a vacuum system are not really interested in the quality of the vacuum itself. They would like to work in a controlled atmosphere and vacuum is the easier and straightforward solution. However, specifying the vacuum system by the process implies to put the responsibility for achieving the desired vacuum quality on the system designer. This is acceptable as long as the system meets the requirements, but could create conflict if the performance does not conform.

A simpler way to define the vacuum quality is to define its requirements based on important operational parameters which should include required pressure to decrease gas scattering and consequently beam lifetime and induced radiation to materials and electronics, pump down time to recover the machine from a sudden intervention or outgassing rate to reduce damage on optical components. The responsibilities to ensure the requested performance are quite clear and rely on the manufacturer of the equipment and of each sub components that should follow strict requirements defined by the vacuum expert. The main factors that need to be taken into consideration and that could affect the vacuum performance are materials choice: mechanical properties, working joints and outgassing rate; cleaning process: when, how and why; required pressure: measurements, pumping speed, vacuum stability; residual gas analysis: partial pressure definition.

In general the lower the required base pressure is, the more rigorous the UHV cleaning is, and the choice of material should need to be. It is important also to stress out that the base pressure refers to the lowest normal pressure attained in a vacuum system in its working condition. Based on this the vacuum expert should determine what level of contaminants can be tolerated and which total and partial pressure of particular species could be accepted.

### 2.2 Design requirements: facts to be considered

It is essential at the design stage to consider methods of the assembly so that components can be successfully individually cleaned. Large equipment should be designed so that it can be cleaned in smaller parts which is much easier to handle. These smaller parts can then be reassembled so that further cleaning is unnecessary.

The most important rule on design UHV components is trying minimizing the gas load while maximizing the effective pumping speed. Minimizing the gas load imply choosing the appropriate surface treatments and minimizing the exposed surface area to the vacuum system. In order to maximize the pumping speed, it is important to carefully design the assembly by maximising the pumping capacity, remove any conductance limitation and place the pump as close as possible to the real outgassing source.

Another important design criterion is the elimination of trapped volumes. Blind holes, crevices and faulty welds must be avoided as they trap dirt and could be filled with cleaning fluids. Blind threaded holes, closed fitting joints and bolted assembly giving trapped volume are some kind of design features that should be avoided. [14]

## 2.3 Compatible materials

In modern accelerator system, stringent requirements are needed for the materials used for vacuum system. Physical and mechanical properties, weldability or brazeability are extremely important parameters that should be addressed [15-18].

By default, only metallic or ceramic materials may be used in the construction of any component or system, which interfaces with the vacuum system. Elastomers or organic materials (Viton, Kapton, PEEK, etc.) are not permitted unless they are specifically authorized for a particular application.

Porous materials such as graphite, ferrite, boron nitride, molybdenum graphite must be analysed and pre-acceptance tests are mandatory to define the adequate cleaning methods and thermal treatments to be followed.

Generally, materials with high vapour pressure are not UHV compatible, in particular cadmium, lead, potassium, sodium, sulphur and zinc and all alloy containing one or more of this materials.

The list below, although not exhaustive, gives a guideline for some materials to be used in UHV applications.

| Type | |
|---|---|
| Pure Metals | Aluminium, Beryllium, Copper, Indium, Molybdenum, Tantalum, Titanium, Tungsten |
| Stainless Steel | Preferred types: 304 L and 316L for beam pipes and 316LN ESR for flanges |
| Alloys | Appropriate aluminium alloys: 6061 grade is good for general use, Beryllium-copper, Glidcop®, Inconel ® |
| Other | Aluminium ceramic, Boron carbide, Diamond sapphire, Macor |

## 2.4 Chemical cleaning for ultra-high vacuum

As a common best practice all components shall be UHV cleaned before assembly. This implies, in particular, no traces of hydrocarbons, organic or inorganic residues from grinding or from the handling and no fingerprints on the inside surface. Adequate protection of all components shall be provided at all times to maintain the state of cleanliness. Sealing materials containing silicon grease are forbidden.

No blind holes of small diameter (less than 3 mm) are permitted in parts to be chemically cleaned for UHV use because chemical solutions may be trapped in these holes. Screw holes should be drilled through or vented to allow the trapped gas to be pumped [19-20].

## 3 Vacuum acceptance requirements

Vacuum acceptance criteria must be defined according to each machine requirements. The main parameter to take into consideration is the admissible outgassing rate for fully assembled components. The total outgassing rate limits should be determined according to machine parameters such as beam lifetime and beam losses function of beam-gas scattering. It could also be defined as a beam downtime described as the allowed time to restart the machine in case of a components exchange. In addition to that, special equipment like injection/extraction kickers or RF cavities could define themselves the maximum admissible pressure level.

In the paragraph below some examples of vacuum requirements used at CERN for unbaked and baked system are presented and the reasoning behind is presented.

### 3.1 Unbaked system

*3.1.1 Pump down curve*

For every vacuum tests performed at CERN for unbaked components, a pump down curve is recorded and analysed. Pressures (P) versus time (t) curves are plotted on a double-logarithmic scale graph. In the high-vacuum region, for metallic substrates, the pressure decrease should be fitted by a 1/t equation:

$$P(t) \propto 1/t^n$$

This equation provides information about the time needed to pump down the vacuum systems. In this condition, water vapour adsorbed onto the surfaces drives the outgassing process of metals in vacuum. By analysing the different slopes, it is possible to have an immediate indication of non-conformity such as the presence of a contaminant and/or virtual leaks. Figure 1 shows an example of 3 curves where the outgassing process is driven by the water vapour adsorbed on the metallic surface ($q_{H2O} \propto 1/t$), by an external air leak and by the water outgassing of a polymeric material ($q_{H2O} \propto 1/\sqrt{t}$) [17].

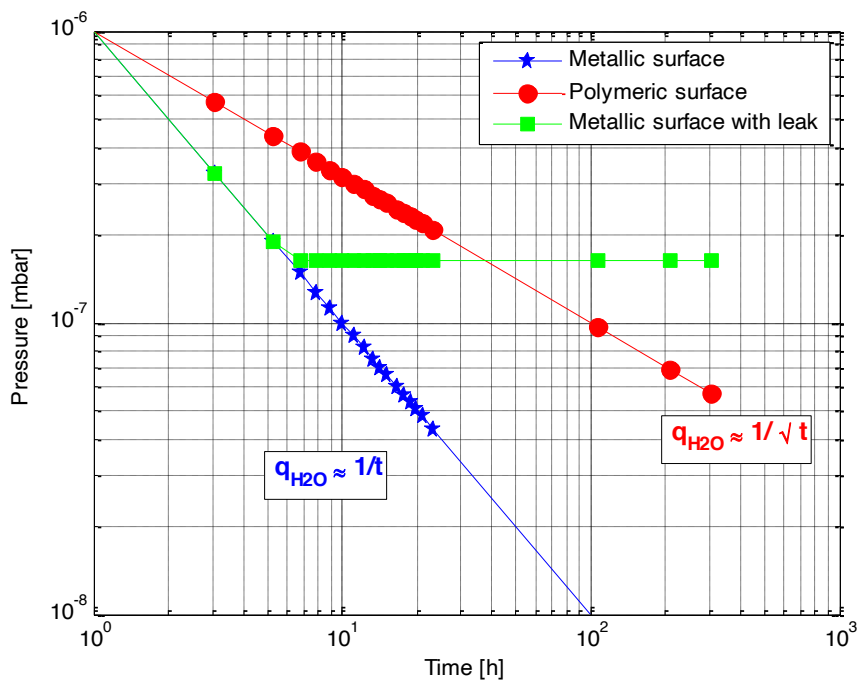

**Fig. 1:** Example of pump down curve for a metallic system, a metallic system with a leak and a component with polymeric materials inside.

*3.1.2 Outgassing rate*

The throughput method is generally used to measure the total $N_2$ equivalent outgassing rate of all tested devices. Such a method is based on the differential pressure measurement by two gauges placed on both sides of a known gas conductance.

The vacuum system of all the injector complex is mainly based on ionic sputter pumps. Following an air or $N_2$ venting of the vacuum sector, a first pump down is always carried out by mobile turbo pumping system. Based on this, the vacuum acceptance criteria is driven by the maximum allowed pressure reached after 24h prior to switching on the ion pumps. The restarting of the ion pumps will allow a subsequent fast pressure decrease allowing the opening of the sector valves of the vacuum sector.

To prevent fast ageing of the ion pumps, for all the injector complex, the maximum pressure after 24h of pump down to start the ion pumps should not be > $5 \cdot 10^{-5}$ mbar. Based on this value and an estimated distributed pumping speed for each accelerators, it was possible to estimate the maximum allowed outgassing rate of each device. Moreover, as a reference, this outgassing rate is compared to the effective outgassing rate of a metallic surface after 24h of pump down that correspond to $\approx 1.0 \cdot 10^{-10}$ mbar·l/s·cm$^2$.

The figure 2 below shows a typical pump down curve for an ARC of the Super Proton Synchrotron (SPS). Different localise outgassing sources are added to simulate the effects of a new device on the pump down time. After 24h, the ion pumps are switched on (pressure drop) which allow shortly after, the opening of the sector valves (P<10$^{-6}$ mbar) and return the machine back to the operation with the beam.

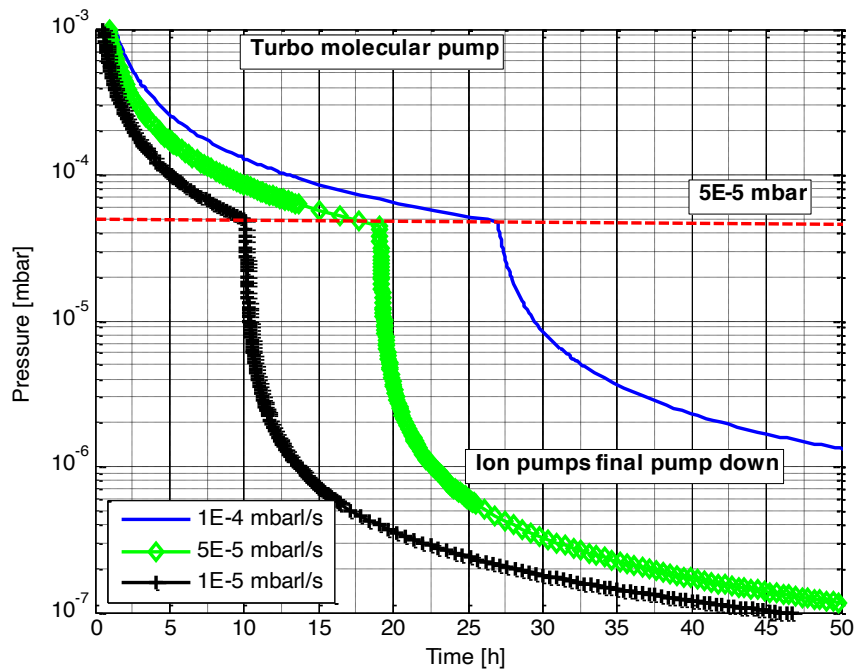

**Fig. 2**: Example of mobile turbo molecular pump down curve followed by sputter ion pump as a function of localised outgassing sources.

### 3.1.3   *Residual gas analysis*

Residual gas analysis (RGA) for each device is performed 24h after the start of the pump down. For an unbaked system, the gas species are normalised to the dominant gas peak: 18 ($H_2O$) atomic mass unit (amu).

Except for the gas species always present in a vacuum system 28 amu (CO) and 44 amu ($CO_2$), for all the mass up to 50 amu two decades lower the $H_2O$ peak is considered. However, for gases with amu higher than 50, a clear signal of contaminants in the system, a limit of 3 decades compared to the $H_2O$ peak is taken into consideration.

As contaminants present in a vacuum system we consider:
- The anomalous presence of hydrocarbons due to:

    - Inadequate cleaning or error in cleaning procedure.

    - Inappropriate choice of materials: polymers, glues, lubricants, etc.

    - Wrong handling of the vacuum chambers: use of gloves and protection must be respected all the time.

- Gases higher than expected CO and $CO_2$ pressures:

    - An indication of carbonised elements.

Figure 3 and 4 shows two RGA scans for conform and not conform components, respectively.

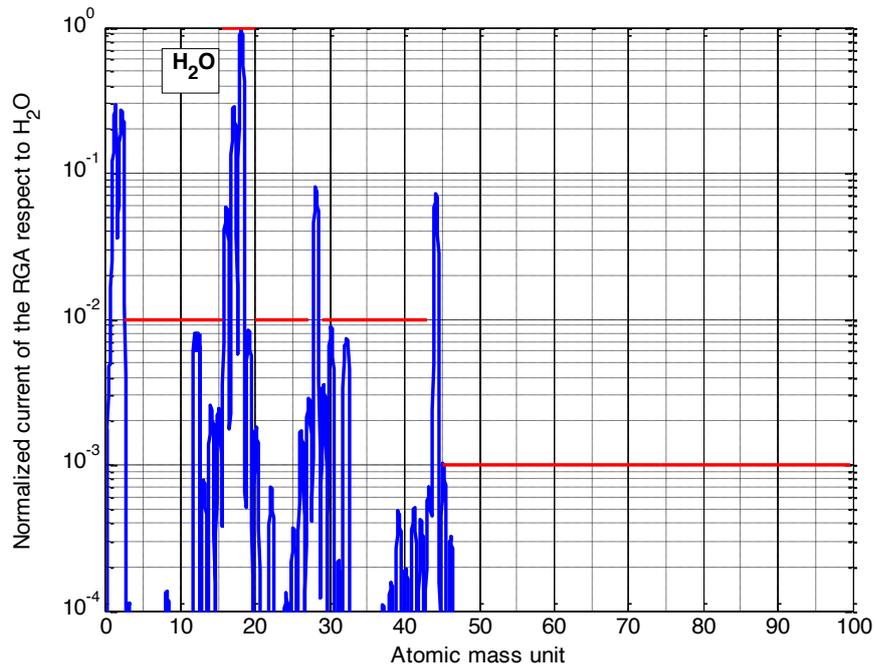

**Fig. 3:** Example of RGA scan of a conforming unbaked system with indicated the acceptance limit

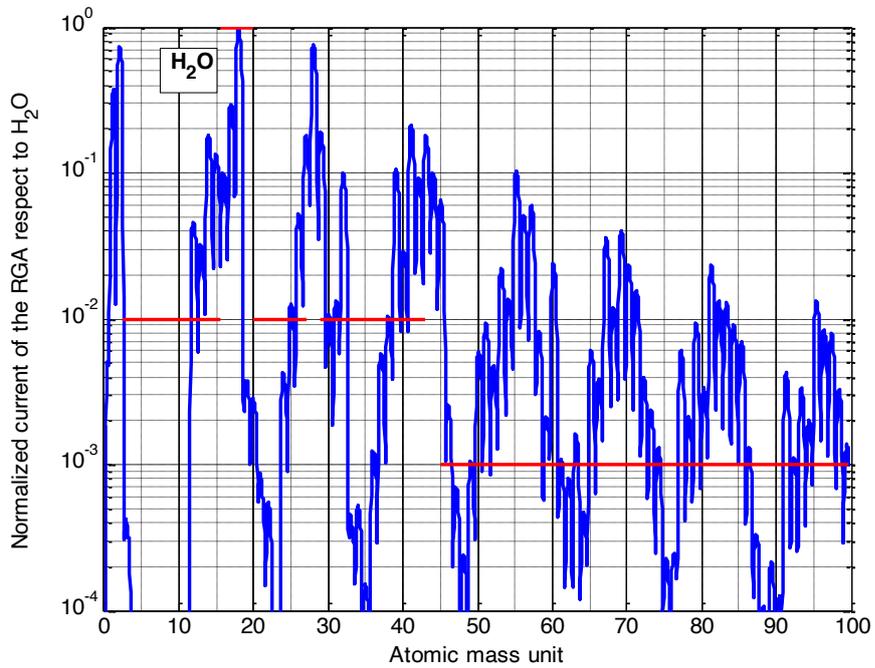

**Fig. 4**: Example of RGA scan of an unbaked system with presence of contaminants.

## 3.2 Baked system: The LHC case

The LHC vacuum requirements were based on the required 100 hours of beam lifetime dominated by nuclear scattering of proton on the residual gas in the beam vacuum chambers. The cross section for such interaction at 7TeV varies with the gas species and are summarized in the table below together with the admissible gas density [21].

**Table 1:** The nuclear scattering cross section at 7TeV for different gases and the corresponding density for 100h of beam lifetime.

| GAS | Nuclear scattering cross section (cm$^2$) | Gas density for a 100h lifetime (m$^{-3}$) |
|---|---|---|
| $H_2$ | $9.50 \cdot 10^{-26}$ | $9.8 \cdot 10^{14}$ |
| He | $1.26 \cdot 10^{-25}$ | $7.4 \cdot 10^{14}$ |
| $CH_4$ | $5.66 \cdot 10^{-25}$ | $1.6 \cdot 10^{14}$ |
| $H_2O$ | $5.65 \cdot 10^{-25}$ | $1.6 \cdot 10^{14}$ |
| CO | $8.54 \cdot 10^{-25}$ | $1.1 \cdot 10^{14}$ |
| $CO_2$ | $1.32 \cdot 10^{-24}$ | $7.0 \cdot 10^{13}$ |

### 3.2.1 Outgassing rate

The throughput method is also used for the baked system, to measure the total $N_2$ equivalent outgassing rate of all tested devices. This is carried out 48 hours after the end of the bake-out cycle; a blank measurement of the system enables verifying the outgassing rate of the measuring system itself.

The acceptance threshold for the outgassing rate is listed in a vacuum specification originally written for the LHC collimators [22]: the total outgassing rate must be lower than $2.0 \cdot 10^{-7}$ mbar·l/s. Considering this outgassing rate and the effective pumping speed available at the location where the specific device is installed, a maximum pressure of $1 \cdot 10^{-8}$ mbar is ensured in the LHC vacuum system that corresponds to the required 100 hours of beam lifetime in stable operation condition as indicated previously.

In addition to that, for small samples the specific outgassing rate should be <$10^{-12}$ mbar·l·s$^{-1}$·cm$^{-2}$ which corresponds to the specific outgassing rate of stainless steel after bake-out at 250˚C without prior thermal treatment (no vacuum firing). In the event that the specific outgassing rate of a component is higher than this acceptance criterion, an evaluation of its impact on the final outgassing rate of the entire components and the achieved pressure level in the LHC, taking in account pressure increase due to dynamic effects and impedance, is carried out.

### 3.2.2 Residual gas analysis

RGA is used to verify the absence of contamination and air leak, and additionally to measure partial pressures after bake-out. In a baked vacuum system, $H_2$ must be the dominant gas. All partial pressures of other gas species have to be lower than that of $H_2$. For this, a template of acceptance limits is defined and applied to RGA currents, normalised to the mass of $H_2$. The template of acceptance limits for baked system is illustrated in Fig. 5. In Fig. 6 an example of a non-conforming device is shown.

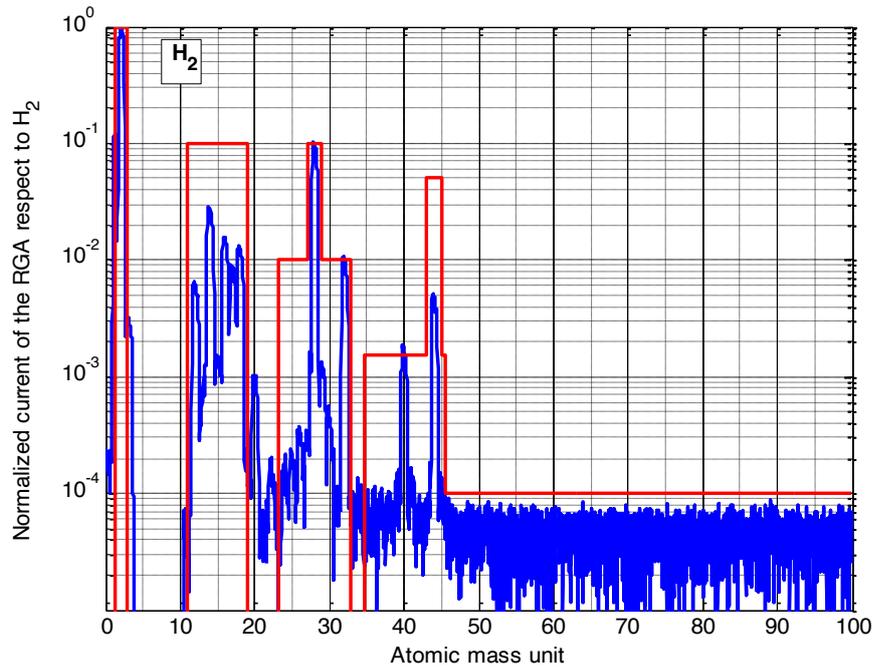

**Fig. 5:** RGA scan normalised to mass peak 2 amu and acceptance thresholds. Collimator after bake-out

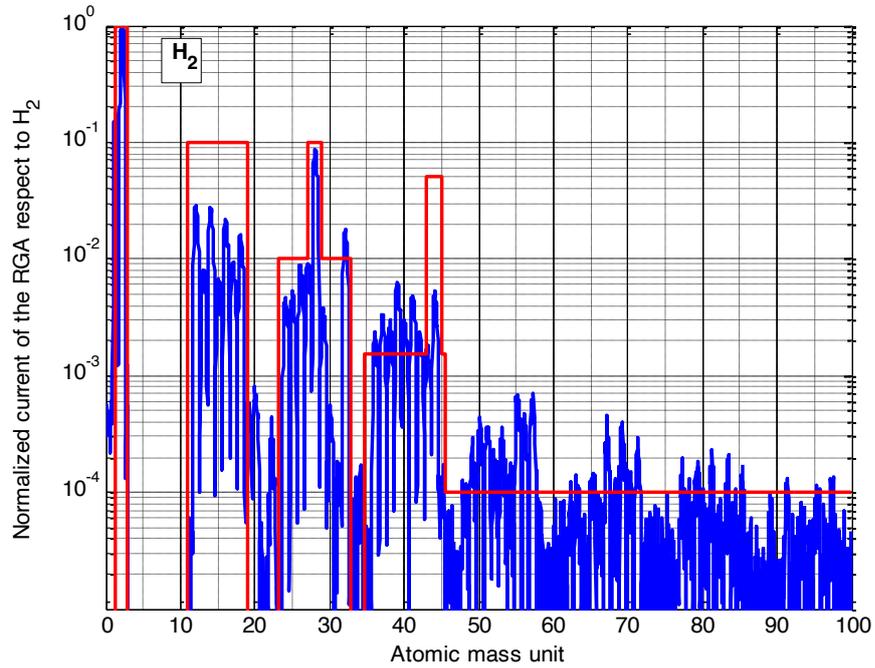

**Fig. 6**: RGA scan normalised to mass peak 2 amu and acceptance thresholds. Non-conforming collimator after bake-out with the presence of contaminant after amu 50

The characteristic mass peaks of $CH_4$, $H_2O$, and CO, i.e. 16, 18, and 28, respectively, have to be at least 10 times lower than the mass peak of $H_2$. Mass 44, the main peak of $CO_2$, has to be a factor of 20 lower than the peak of $H_2$. Finally, heavy-hydrocarbon with mass peaks over 50 amu, if detected, must have intensities 10000 times smaller than mass peak of $H_2$.

### 3.2.3 Virtual leaks

After bake-out, no signal of argon must be recorded by the RGA. When the RGA scan shows typical air mass peaks (Figure 7: $N_2^+$ 28 amu, $N^+$ 14 amu and $Ar^+$ 40 amu) and no external leak is detected by the use of a helium leak detector, the air virtual leak rate is estimated by an accumulation test. Such a test is performed by isolating the test system from all active pumping. Pressure and RGA signals are monitored during the accumulation time as shown in Fig. 8.

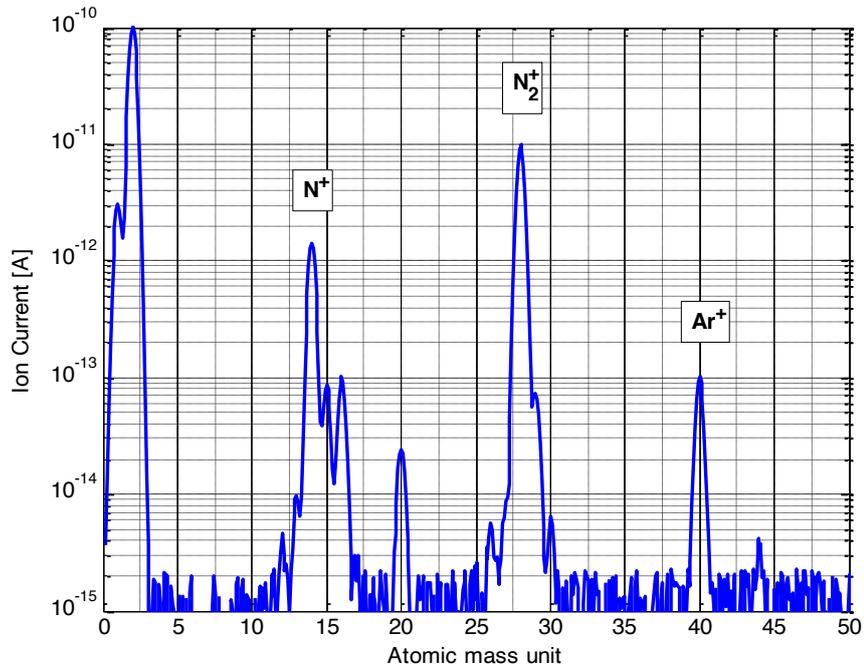

**Fig. 7**: RGA scan with a clear indication of air leak. Atomic mass of 40 ($Ar^+$), 28 ($N_2^+$) and 14 ($N^+$) indicating the presence of a leak.

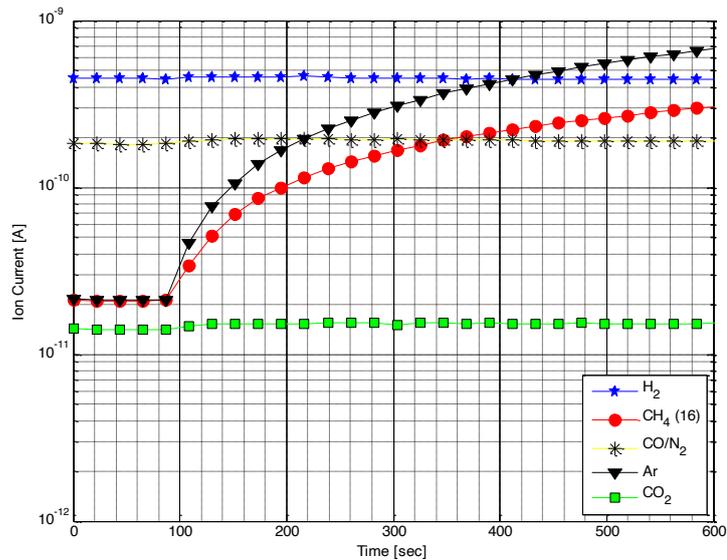

**Fig. 8**: Example of accumulation test of tested components. In this case, the active pumping of NEG after bake-out for all gases except $CH_4$ and noble gases, enables the detection of internal leak in the system.

The virtual leak rate is estimated by calculating the increase of the Ar peak signal over a fixed period of time. The RGA calibration curves are then used to estimate the Ar partial pressure and consequently the total air internal leak. The acceptance limit for a virtual air equivalent leak in the LHC UHV system is fixed to less than $5.0 \cdot 10^{-9}$ mbar·l/s. For this value, the virtual leaks would produce at most the complete saturation of 1 m long, 8 cm diameter NEG-coated standard beam pipe in about 150 days.

## 4  Summary

This paper has briefly outlined some basic requirements to be put in place for vacuum acceptance tests of components to be installed in different vacuum accelerators. It draws attention to the need of defining proper and clear acceptance criteria that enable both safe operation and intervention in the different vacuum systems and guarantee the integrity of the vacuum system over the years.

As the installation of the vacuum system is one of the last steps of the accelerator installation, the available time is unfortunately often reduced. It is important to bear in mind that the vacuum acceptance tests are essential parts of the overall work in order to deliver a vacuum system that fulfils the accelerator requirements. Different steps need to be carried out ranging from the pump-down curve, leak detection, total outgassing, partial pressure and residual gas analysis and possible internal leak rate.

## 5  Acknowledgements

I would like to thank my colleagues of the Vacuum, Surfaces and Coatings (VSC) group at CERN for their support and the many fruitful discussions.